\documentclass[aps,prd,nofootinbib,longbibliography,twocolumn]{revtex4-1}

\usepackage{amsmath}
\usepackage{amssymb} 
\usepackage[cm]{fullpage}
\usepackage{graphicx} 
\usepackage{siunitx}
\usepackage{tabularx} 
\usepackage[colorlinks,linkcolor=blue,citecolor=blue,urlcolor=blue]{hyperref}
\usepackage[utf8]{inputenc}
\usepackage{hyperref}
\usepackage{cleveref}

\DeclareSIUnit\parsec{pc}

\newcommand{\be}{\begin{equation}}
\newcommand{\ee}{\end{equation}}
\newcommand{\beqx}{\begin{equation*}}
\newcommand{\eeqx}{\end{equation*}} 
\newcommand{\beqa}{\begin{eqnarray}}
\newcommand{\eeqa}{\end{eqnarray}}
\newcommand{\beqax}{\begin{eqnarray*}}
\newcommand{\eeqax}{\end{eqnarray*}}

\begin{document}

\title{Dark Energy, the Swampland and the Equivalence Principle}
\date{\today}
\author{Carsten van de Bruck and Cameron C. Thomas}

\affiliation{Consortium for Fundamental Physics, School of Mathematics and Statistics, University of Sheffield, Hounsfield Road, Sheffield S3 7RH, United Kingdom}

\begin{abstract}
It has recently been argued that string theory does not admit de Sitter vacua. This would imply that the current accelerated expansion of the universe is not driven by a cosmological constant (or vacuum energy) but by other means such as a quintessential scalar field. Such a scalar field is in general expected to couple to at least some matter species, such as dark matter. Cosmological observations already constrain such dark matter couplings strongly. We argue that there are a number of interesting scenarios to be explored, such as coupling functions which possess a minimum at finite field values. In these theories, the effective gravitational coupling between dark matter particles grows with time and are consistent with observations of the anisotropies in the cosmic microwave background radiation and large scale structures. We argue that such couplings might also help to alleviate the tension between the swampland conjectures and the properties of the quintessential potential. Observational signatures of violations of the equivalence principle in the dark sector are expected in the non-linear regime on intermediate or small scales. 
\end{abstract}

\maketitle

\section{Introduction}
It is well known that the current accelerated expansion of the universe, described very successfully by the cold dark matter (DM) model with a cosmological constant ($\Lambda$CDM), 
is difficult to construct from the bottom--up in particle physics models beyond the standard model. More dramatically, 
it has recently been conjectured that de Sitter space cannot be embedded in string theory and that there are restrictions on the effective field 
theory compatible with string theory \cite{Obied:2018sgi}. These restrictions are called the swampland criteria and are given below. 
While this conjecture is currently under scrutiny, if true it would either imply that string theory, as currently understood, is wrong or that the current accelerated expansion is not due to a non-vanishing cosmological constant (or vacuum energy) but is driven by some other processes, either by modifications of gravity or new degrees of freedom in the matter sector (dark energy (DE)). Among the best studied 
models are quintessence fields, in which the expansion at late times is driven by a light scalar field \cite{Wetterich:1987fm,Ratra:1987rm,Caldwell:1997ii,Zlatev:1998tr}. The swampland criteria on the effective potential $V$ of scalar fields are given by the following inequalities:\footnote{Extensions of these criteria have been proposed in \cite{Andriot:2018wzk,Ooguri:2018wrx,Andriot:2018mav}.}
\be \label{swampland1}
 M_{\rm Pl}|\nabla V| \geq c V,
\ee
where the left hand side denotes the gradient of the potential, $M_{\rm Pl}$ is the reduced Planck mass and $c\sim {\cal O}(1)$ is a constant. In addition, the field should not vary more than 
roughly one Planck unit throughout the history of the universe
\be\label{swampland2}
\Delta\phi \lesssim d~ M_{\rm Pl}~,
\ee
where $d \sim {\cal O}(1)$, as otherwise light fields become important and the effective field theory becomes invalid (see \cite{Palti:2019pca} for a recent review). The swampland criteria have sparked a lot of activity recently, because of their implications for inflation (see e.g. \cite{Banerjee:2018qey,Achucarro:2018vey,Garg:2018reu, Kehagias:2018uem,Brahma:2018hrd,Dimopoulos:2018upl,Kinney:2018nny,Kinney:2018kew,Lin:2018kjm}) and dark energy (see e.g. \cite{Agrawal:2018own,Chiang:2018jdg,Heisenberg:2018yae,Cicoli:2018kdo,Akrami:2018ylq,Heisenberg:2018rdu,Marsh:2018kub,DAmico:2018mnx,Han:2018yrk,Heckman:2018mxl,Olguin-Tejo:2018pfq,Colgain:2018wgk,Heckman:2019dsj,Brahma:2019kch}).

It has long been argued that a quintessence field should couple to other sectors, 
unless there is a symmetry which forbids this \cite{Carroll:1998zi}. This is certainly the case for quintessence fields motivated from string theory, in which the fields usually determine coupling constants. In \cite{Agrawal:2018own}, the authors argued that the field should at least couple to dark matter, given that the couplings to the standard model particles are strongly constrained. Theories with DE-DM couplings have been studied for some time and models like these are named 'coupled quintessence' \cite{Wetterich:1994bg,Amendola:1999er}. These type of theories are already strongly constrained by current observations of the cosmic microwave background, large scale structures and the expansion history of the universe. As we will see below, for the simplest theories, the couplings allowed are very small. The strength of a fifth force mediated between two DM particles is constrained to be much less than that of gravity \cite{vandeBruck:2017idm,Miranda:2017rdk}, if the coupling is constant. In this paper we argue that one class of models, which could be very relevant for the swampland picture, deserves further studies. In these models the fifth force between DM particles switches on at late times, when the quintessence field starts to roll down its potential energy. These theories can be reconciled with observations of the cosmic microwave background radiation (CMB) as well as large scale structures (LSS), but they predict an equation of state different from the cosmological constant (and in fact can mimic DE with an equation of state $w\lesssim -1$) and an effective gravitational constant between DM particles considerably larger than Newton's constant $G_N$. We expect the predictions for structure formation and evolution in these models to deviate from $\Lambda$CDM model on intermediate and smaller length scales. 

The paper is organized as follows: in the next Section we study first the simplest types of models, based on a conformal coupling. We then modify the theory with a different coupling function which possesses a minimum. In Section 3 we further extend the class of models, allowing for derivative couplings. Our conclusions and an outlook are given in Section 4. 

\section{The simplest models and an extension}
A well studied effective field theory setup to introduce a coupling between dark matter and a scalar field $\phi$, is the following. The gravitational field and the standard model field propagate under the influence of the 
metric $g_{\mu\nu}$, whereas dark matter (DM) particles propagate on geodesic with respect to a second metric ${\tilde g}_{\mu\nu}$. The gravitational field and the scalar field are described by the action

\be
{\cal S} = \int d^4 x \sqrt{-g}\left( \frac{M_{\rm Pl}^2}{2} {\cal R} - \frac{1}{2}g^{\mu\nu} \partial_\mu \phi \partial_\nu\phi - V(\phi) \right) ~,
\ee
where ${\cal R}$ is the Ricci--scalar and $V(\phi)$ the potential energy of $\phi$. The action for the matter fields consists of 
\be
{\cal S}_{\rm SM} = \int d^4 x \sqrt{-g} {\cal L}_{\rm SM}(g,\Psi_i)
\ee
for the standard model fields $\Psi_i$ and 
\be \label{DMaction}
{\cal S}_{\rm DM} = \int d^4 x \sqrt{-{\tilde g}} {\cal L}_{\rm DM}({\tilde g},\sigma)
\ee
for the DM field $\sigma.$ \footnote{We assume here for simplicity that there is only one dark matter species. This doesn't have to be the case and in \cite{Brookfield:2007au,Baldi:2012kt} examples of more complicated setups have been studied.} In what follows, we are interested in processes well after Big Bang Nucleosynthesis (BBN) and assume that dark matter consists of non--relativistic particles. 
In the simplest case, the relationship between the two metrics is via a conformal transformation ${\tilde g}_{\mu\nu} = C(\phi) g_{\mu\nu}$, where $C(\phi)$ is a generic function of $\phi$. As a consequence, 
DM particles do not propagate on geodesics with respect to the metric $g_{\mu\nu}$ but there is an additional force, mediated by $\phi$ acting on DM particles. For long--ranged forces, which in case of 
the light quintessence fields we are considering here are of order of the horizon size, the effective Newton's constant is given by ($G_N$ is Newton's constant)
\be\label{effgrav}
G_{{\rm eff}} = G_N\left( 1 + 2\beta^2 \right)~,
\ee
with 
\be \label{coupling}
\beta \equiv \frac{M_{\rm Pl}}{2} \left(\frac{{\rm d} \ln C}{{\rm d} \phi} \right)
\ee
is the coupling. From the perspective of the action, the 
masses of the DM particles become field dependent with $m(\phi) = m_0 \sqrt{C(\phi)}$, where $m_0$ is a bare mass parameter setting the overall mass scale of the DM particles (see e.g. \cite{Fujii:2003pa}). This exchange of dark energy (DE) 
and non--relativistic DM implies that the DM density does not scale like baryons (for which the energy density $\rho_b$ scales like $a^{-3}$, where $a$ is the scale factor describing the expansion of space), but depending on 
the details of the energy exchange decays either faster or slower than baryons. The density of DM particles is given by $\rho_c = m(\phi)/a^3$, resulting in the energy conservation \cite{Wetterich:1994bg} 
\be
\dot \rho_c + 3 H \rho_c = \frac{1}{2} \frac{\rm d \ln C}{\rm d \phi} \dot\phi \rho_c = M_{\rm Pl}^{-1} \beta \dot\phi \rho_c,
\ee
where $H = \dot a/a$ is the expansion rate and the dot denotes the derivative with respect to cosmic time. This equation can be also obtained directly from the action (\ref{DMaction}), where the energy momentum tensor for dark matter is defined as 
\be
T_{\mu\nu} = -\frac{2}{\sqrt{-g}} \frac{\delta {\cal S}_{\rm DM}}{\delta g^{\mu\nu}}, 
\ee
and DM is modelled as a non--relativistic fluid. The evolution of the scalar field is given by

\be
\ddot \phi + 3H\dot\phi + V_{,\phi} = - M_{\rm Pl}^{-1}\beta \rho_c.
\ee

The modified energy conservation equation implies that, in general, the ratio of the density parameter of baryons and DM $\Omega_b/\Omega_{\rm DM} = \rho_b/\rho_{\rm DM}$ at last scattering is different as it would be expected in the $\Lambda$CDM model. As a consequence, in these theories the  
positions and relative heights of the peaks and valleys in the CMB anisotropy power spectrum have changed compared to the $\Lambda$CDM 
model \cite{Mifsud:2017fsy}. 

For concreteness, in this paper we will focus on the case of an exponential potential for the scalar field with $V(\phi) = V_0 \exp (-\lambda\phi/M_{\rm Pl})$. A widely studied example for the coupling is the case of ($\alpha$ is constant)
\be \label{coupledquintessence}
C(\phi) = \exp(2\alpha\phi/M_{\rm Pl}),
\ee
in which the coupling (\ref{coupling}) is given by $\beta=\alpha$ and is constant \cite{Wetterich:1994bg,Amendola:1999er}. For this particular choice, the effective gravitational coupling between two DM particles is enhanced by a constant factor $1 + 2 \alpha^2$. 

Constraints coming from supernovae data, baryonic acoustic oscillations and large scale structures put  
upper bounds on the parameter $\alpha$ and $\lambda$, which are roughly $\alpha \lesssim 0.05$ and $\lambda \lesssim 1$ at 2$\sigma$ \cite{vandeBruck:2017idm}. That means that the additional 
force has to be $5\times 10^{-3}$ smaller than gravity in this model. Given that $\alpha$ is constrained to be very small, this model looks rather unattractive. 

\subsubsection{Dark energy and the least coupling principle}

While the coupling function (\ref{coupledquintessence}) is well motivated, we will consider an appealing modification, in which the coupling function $C(\phi)$ possesses a minimum. Since according to the swampland conjectures the evolution of the scalar field is below the Planck mass (see eq. (\ref{swampland2})), we expand $C(\phi)$ around its minimum at $\phi_*$:
\be \label{conformalgrowing}
C(\phi) \approx 1 + \frac{1}{2M_{\rm Pl}^2}\alpha (\phi - \phi_*)^2 + {\cal O} (( \phi - \phi_* )^3).
\ee
In this paper we choose, without the loss of generality, $\phi_* = 1 M_{\rm Pl}$. Furthermore, we will choose $\alpha$ to be positive, so that $\phi_*$ is the minimum of the coupling function and not the maximum. Most of our calculations below assume $\phi_{\rm ini} = \phi_*$, but we will briefly discuss the effect of a displacement from $\phi_*$ in the very early universe as well as discuss two processes in the very early universe which drive the field towards $\phi_*$. The case for coupling functions with minima in string theory has been discussed in \cite{Damour:1994zq,Damour:1994ya} and in \cite{Brax:2010gi} in the context of dark energy physics. 

Since the density of DM does not scale like $a^{-3}$, we define an effective energy density for DE as $\rho_{DE} = \rho_\phi + \rho_c - \rho_{c,0}a^{-3}$. We have therefore split the part of $\rho_c$, which does not scale $a^{-3}$ and included it in the energy density for the DE sector. As a result, the effective equation of state for DE is now given by \cite{Das:2005yj}

\begin{figure}{
\includegraphics[scale=0.16]{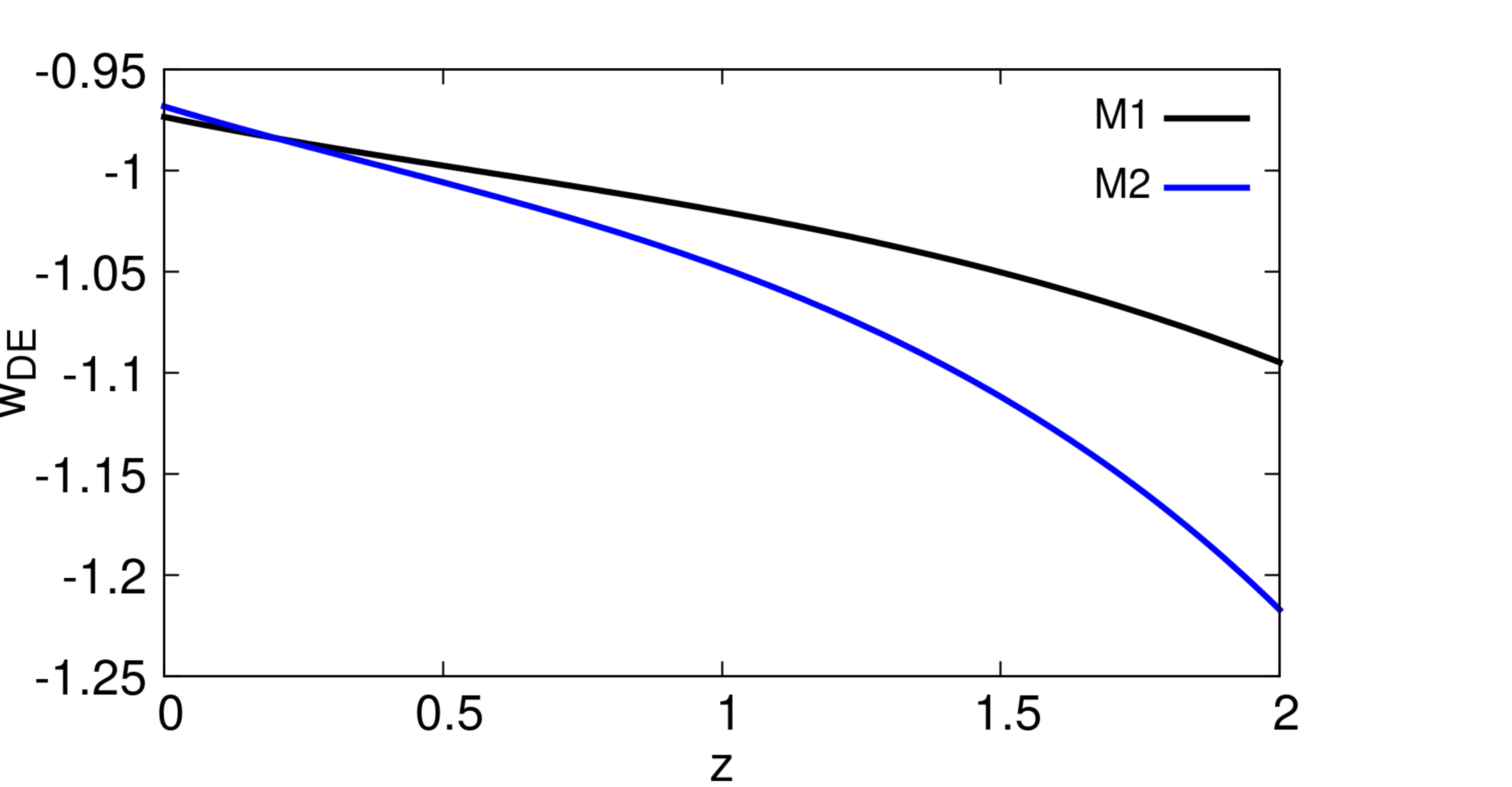}
\includegraphics[scale=0.16]{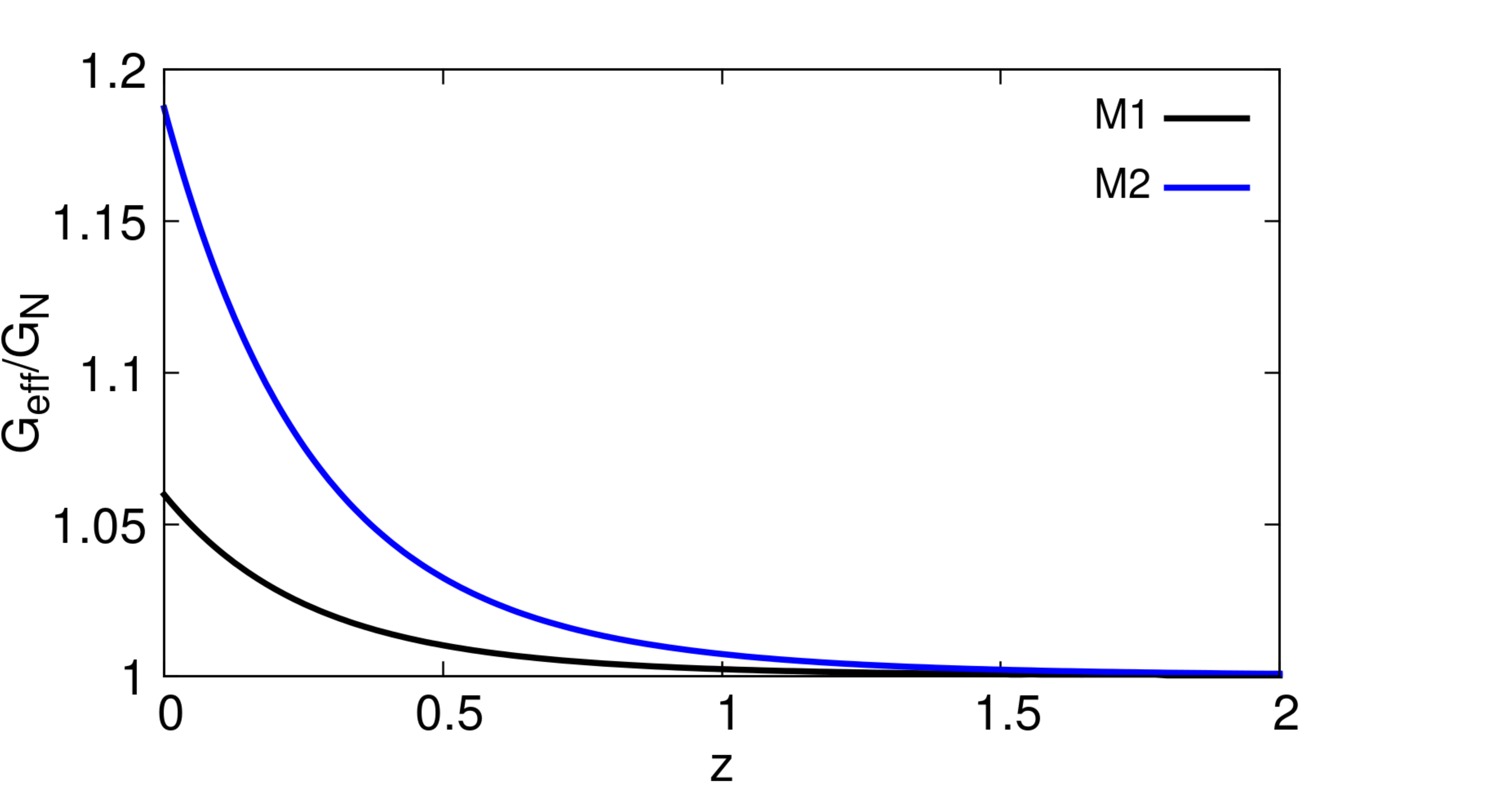}
\caption{Upper figure: Evolution of the equation of state of dark energy $w_{\rm DE}$, defined in Eq. (\ref{weff}), for models M1 ($\alpha = 3, \lambda = 0.5$) and M2 ($\alpha = 5, \lambda = 0.6$). Lower figure: The evolution of the effective gravitational constant, defined in Eq. (\ref{effgrav}), in both models.}
}
\end{figure}

\be \label{weff}
w_{\rm DE} = \frac{p_\phi}{\rho_{\rm DE}} = \frac{p_\phi}{\rho_\phi + \rho_c - \rho_{c,0}a^{-3}},
\ee
where $\rho_\phi$ and $p_\phi$ are the energy density and pressure of the scalar field, respectively. From the definition, $w_{\rm DE,0} = \left(p_\phi/ \rho_\phi\right)_0=w_{\phi,0}$ today in all models. In the uncoupled case, this expression becomes the standard expression $w_{\rm DE} = w_\phi = p_\phi/ \rho_\phi$ for all times. 

To study numerically the background evolution as well as the evolution of cosmological perturbations, we have used a modified version of the CLASS code \cite{Blas:2011rf,vandeBruck:2017idm}. We refer to \cite{Mifsud:2017fsy} and references therein, where all relevant equations for the cosmological perturbations can be found. In the following, we will consider two parameter choices. The first model (M1) has the parameters $\alpha = 3$ and $\lambda = 0.5$, whereas in the second model (M2), we choose $\alpha = 5.0$ and $\lambda = 0.6$. And to be concrete in the examples, we fix the other cosmological parameter as follows ($h$ is the Hubble parameter today in units of 100 km s$^{-1}$Mpc$^{-1}$): Hubble parameter today $H_0 = 67.32$ km s$^{-1}$Mpc$^{-1}$, density parameter of baryons $\Omega_b h^2 = 0.022383$, density parameter of dark matter $\Omega_c h^2 = 0.12011$, spectral index $n_s = 0.96605$, scalar amplitude $\ln(10^{10} A_s) = 3.0448$, optical depth $\tau = 0.0543$. The evolution of the effective equation of state $w_{\rm DE}$ for the scalar field is shown in the upper plot in Figure 1. In both models, the evolution of the scalar field is such that the variation $\Delta\phi <1 M_{\rm Pl}$. 

\begin{figure}
\includegraphics[scale=0.16]{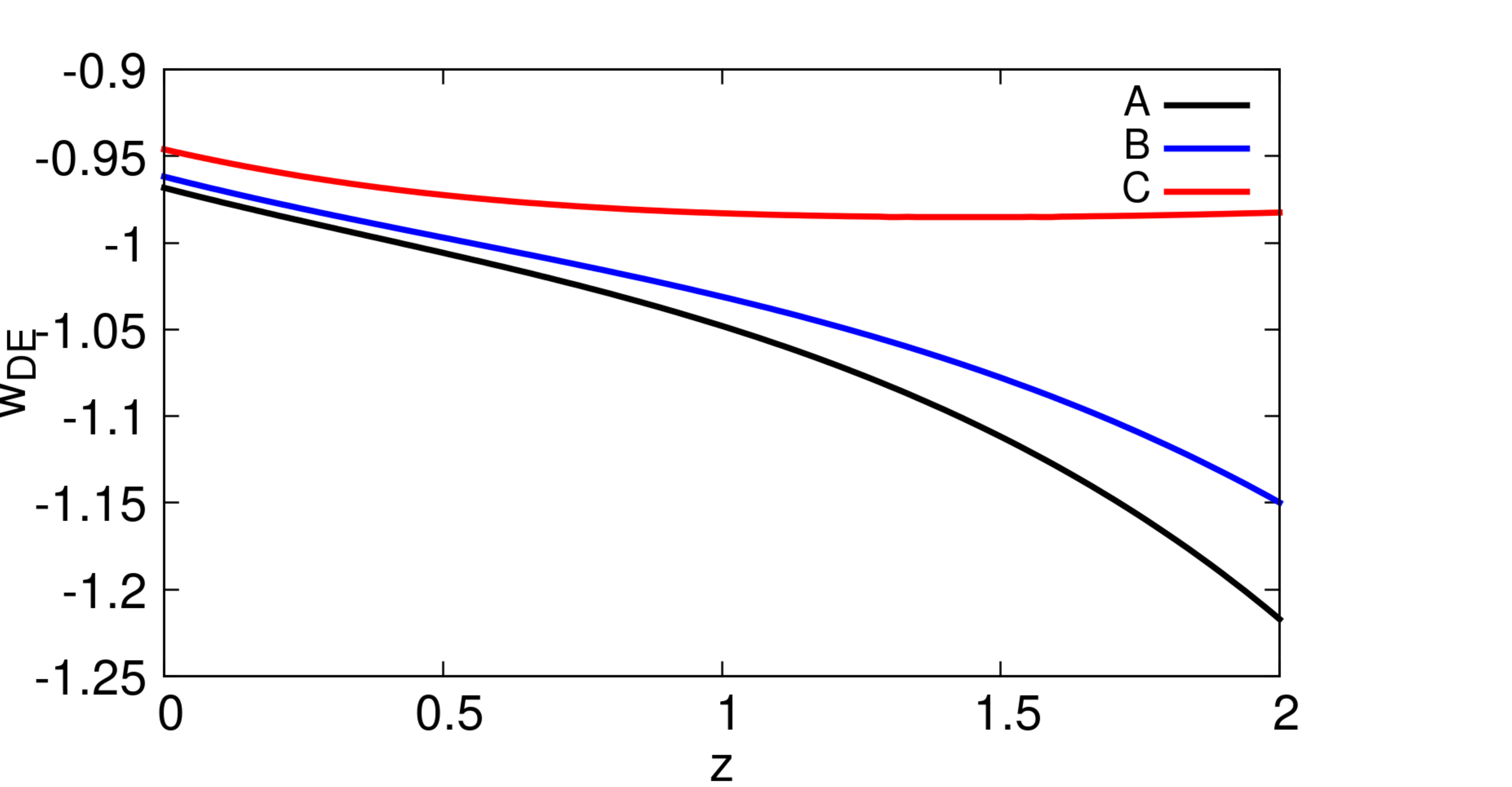}
\caption{Predictions for the equation of state $w_{\rm DE}$ for fixed value of ($\lambda = 0.6$) but different values for $\alpha$: 
A: $\alpha = 5$, B: $\alpha = 3$ and C: $\alpha = 0$ (uncoupled).}
\end{figure}

The evolution of the effective gravitational constant in both models is shown in the lower plot in Figure 1. As it can be seen from this figure, once the potential energy of the scalar field becomes dynamically important, the field starts to evolve and the coupling between DM and DE increases. As a result, a long--range fifth force between DM appears at a redshift of about $z\approx 2$. In both models, the gravitational coupling between DM particles today is considerably larger than $G_N$. 

Note that the same value of $\lambda$ does lead to different predictions of the equation of state today if $\alpha$ is varied (see Figure 2). This is because while the field wants to roll down the potential to larger values of $\phi$, its motion is hindered by the presence of the coupling, which grows as $\phi$ gets larger. Thus, unlike in the uncoupled counterpart, the model discussed here may allow for slightly larger values of $\lambda$. This strongly suggests that the couplings to DM thus help to alleviate the tension of the model with the swampland conjecture (\ref{swampland1}). In table I, we show the results for $\omega_{\rm eff,0}$, $\sigma_8$ and $G_{\rm eff,0}/G_{N}$ for different choices of $\alpha$ but fixed $\lambda = 0.6$. As it can be seen, the effective equation of state moves indeed towards $\omega = -1$, but for very large values of $\alpha$, the value of $\sigma_8$ becomes larger. We will in future work compare the theory to cosmological data, but it is clear that large violations of the equivalence principle in the dark sector today are not necessarily disallowed by current cosmological observations. 

\begin{table}[t] 
   \label{tab:example}
   \small % text size of t%dable content
   \centering % center the table
   \begin{tabular}{c c c c} % alignment of each column data
   \hline\hline
   \textbf{$\alpha$} & \textbf{$\omega_{\rm DE,0}$} & \textbf{$\sigma_{8}$} & \textbf{$G_{\rm eff,0}/G_{N}$} \\ 
   \hline  
   0  & -0.947 & 0.805 & 1 \\
   3  & -0.962 & 0.799 & 1.09\\
   5  & -0.969 & 0.799 & 1.19\\
   10  & -0.979 & 0.801 & 1.46\\
   50 & -0.995 & 0.821 & 2.91\\
   \hline\hline
   \end{tabular}
   \caption{Values of the effective equation of state today ($\omega_{\rm DE,0}$), $\sigma_8$ and $G_{\rm eff,0}/G_{N}$ today for fixed values of $\lambda= 0.6$ but different values of the coupling parameter $\alpha$. As it can be seen, the effective equation of state today approaches the cosmological constant value $\omega = -1$ for larger values of $\alpha$. The value of $\sigma_8$  is large for large values of $\alpha$, implying a stronger clustering of matter.}
  
\end{table}

The predictions for the CMB anisotropies spectrum are shown in Figure 3, in which we compare models M1 and M2 to the $\Lambda$CDM model. As it can be seen, both models M1 and M2 deviate only slightly from $\Lambda$CDM, despite the effective gravitational coupling today being considerably larger. We will constrain the model with cosmological data in future work. Furthermore, it would be interesting to consider changes to the form of the potential, e.g. considering a double exponential \cite{Barreiro:1999zs} or changing the kinetic term of the scalar field (e.g. see \cite{Teixeira:2019tfi}).

\begin{figure}
\includegraphics[scale=0.16]{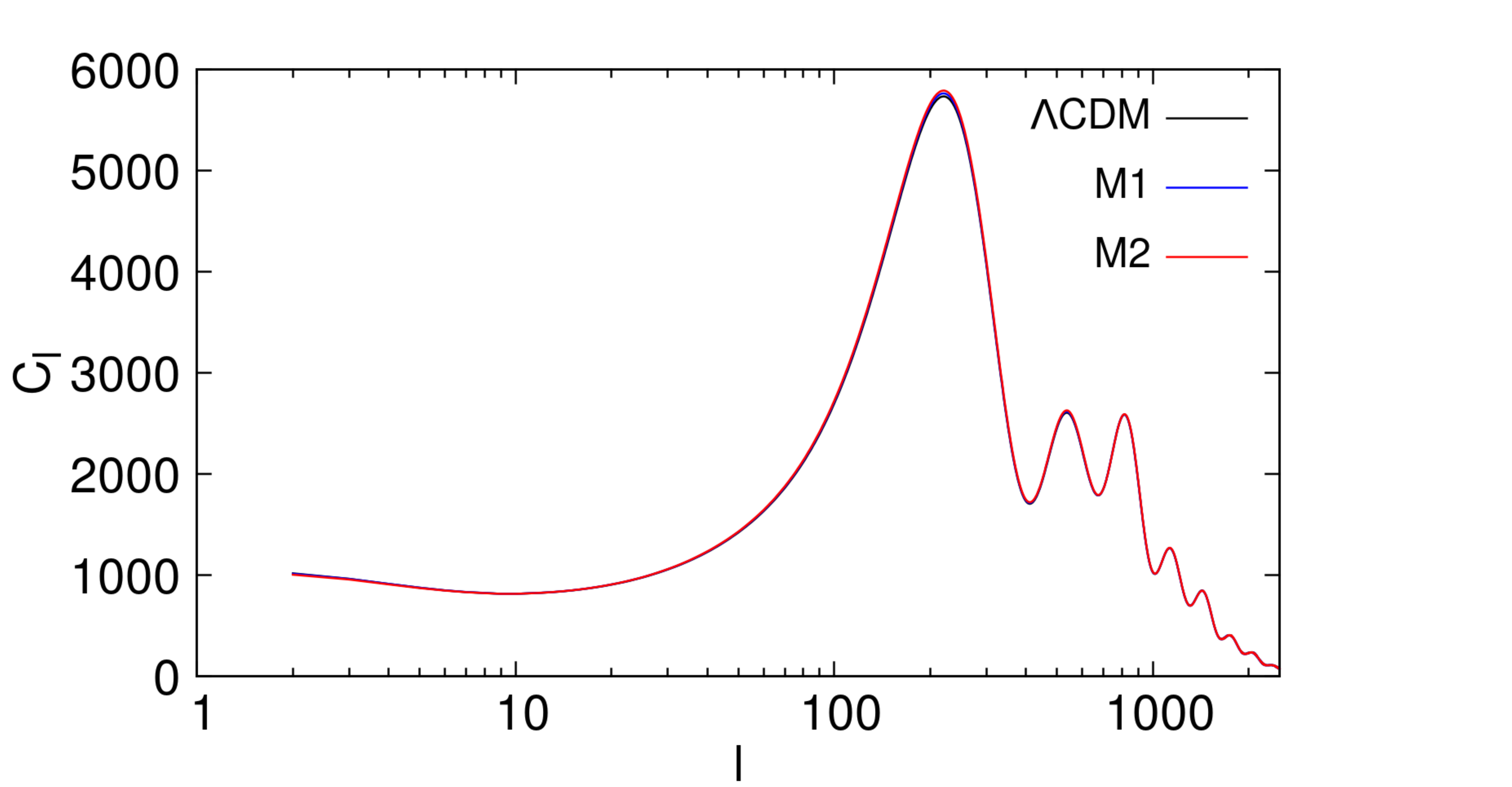}
\caption{The CMB anisotropy power spectrum for models M1 and M2, compared to the $\Lambda$CDM model.}
\end{figure}

\begin{figure}
\includegraphics[scale=0.16]{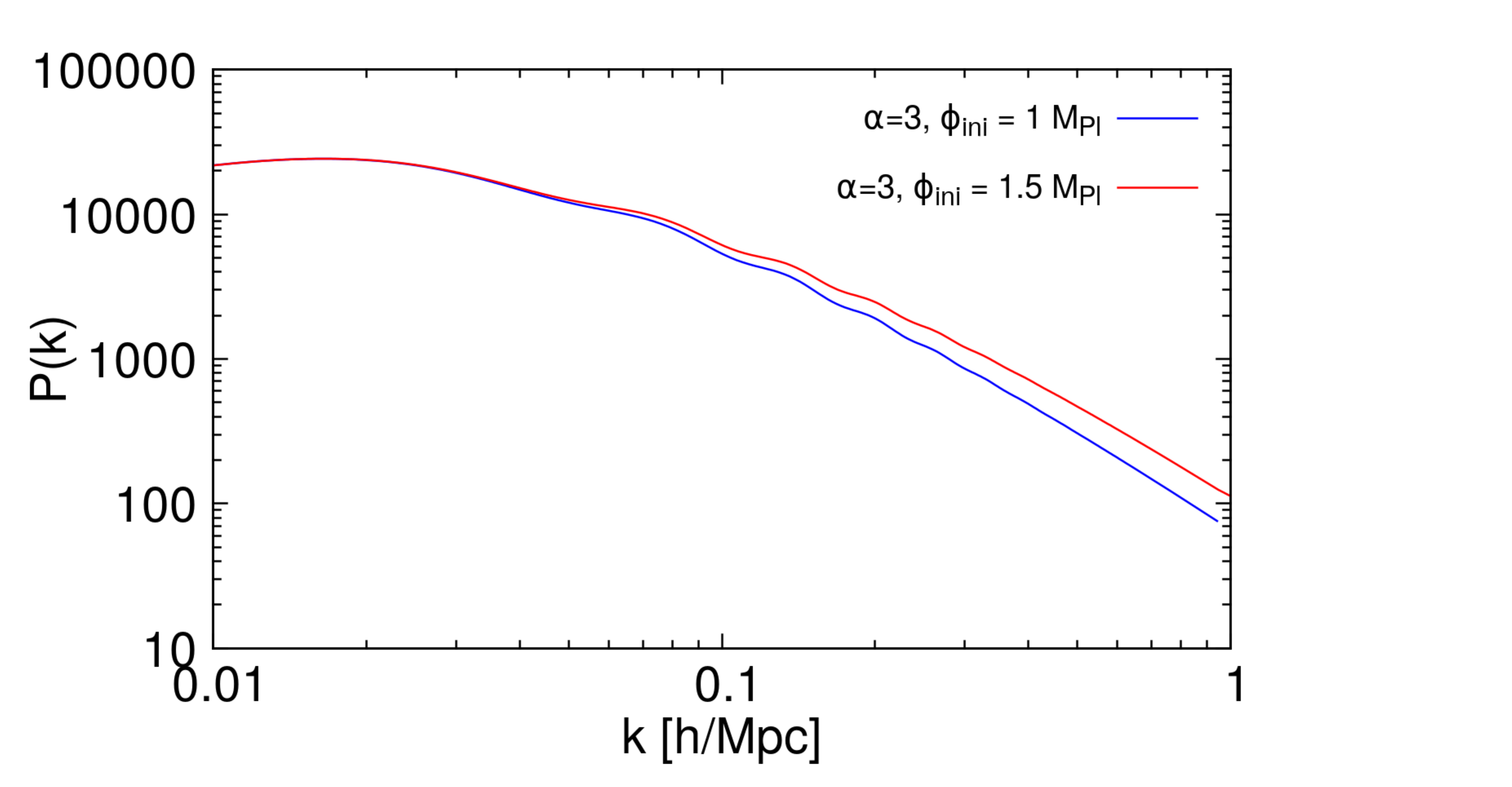}
\includegraphics[scale=0.16]{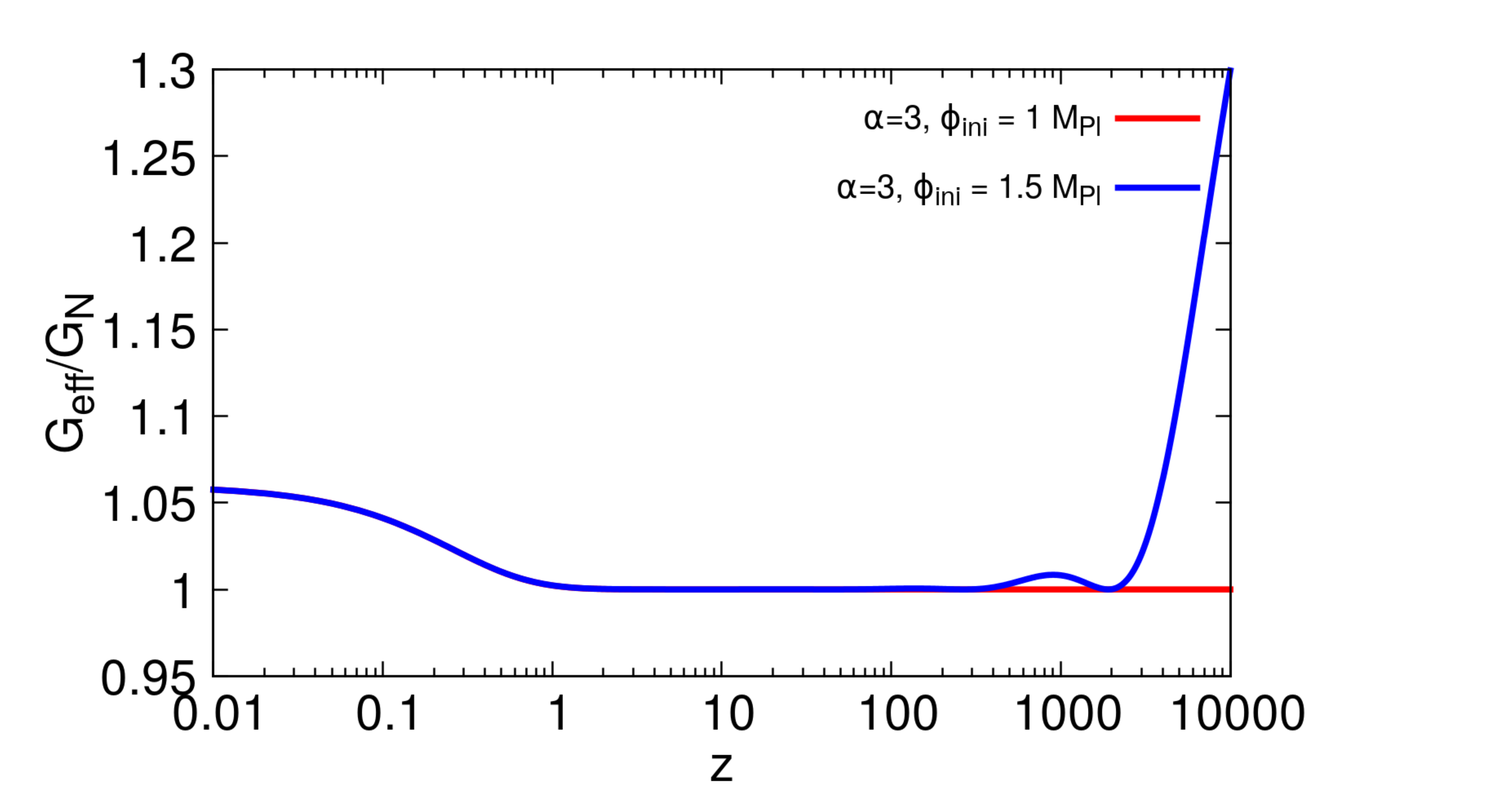}
\caption{Upper plot: The predictions for the linear matter power spectrum for different initial conditions for the scalar field. For both curves, we have chosen $\lambda = 0.5$ and $\alpha = 3$. Lower plot: Evolution of the effective gravitational constant $G_{\rm eff}$ for different initial conditions for the scalar field. The evolution of $G_{\rm eff}$ is essentially the same in both models for redshifts below $z=200$, but differs substantially at higher redshifts.}
\end{figure}

We end this section by addressing the initial conditions for the DE field. The model seems fine--tuned, in the sense that we have chosen the value of $\phi$ in the very early universe to be at the extremum of the coupling function $C(\phi)$. If the field would not be at the minimum $\phi_*$ in the very early universe, the effective gravitational constant would be larger than $G_N$ early on. In this case, we find that the matter power spectrum is enhanced for small and intermediate wavenumbers $k$, as shown in Fig 4, where we fix $\lambda$ and $\alpha$, but vary $\phi_{\rm ini}$. We also show the evolution of the effective gravitational constant. The field starts at a value $\phi_{\rm ini}$ away from the minimum. As soon as the matter density becomes important, the coupling affects the evolution of the field and drives it towards the minimum $\phi_*$. Once it has settled at the minimum, the evolution is the same as in a model for which $\phi_{\rm ini} = \phi_*$. The enhanced gravitational constant in the late radiation dominated epoch results in an enhancement of the power spectrum and therefore larger values for $\sigma_8$. 

\subsubsection{Attractor mechanism in the very early universe}

There are two processes which drive the field towards the minimum of the coupling function $C(\phi)$ in the very early universe. One mechanism is provided by inflation and the other one at the time dark matter becomes non--relativistic. 

Let us first consider a set of inflaton fields $\chi_i$, which roll down their potential energy $U(\chi_i)$. If $\phi$ is coupled to the fields $\chi_i$ in the same way as it is coupled to DM, then the action of the theory is
\begin{eqnarray}\label{inflaaction}
{\cal S} = \int d^4 x \sqrt{-g} \left[ \frac{M_{\rm Pl}^2}{2} R - \frac{1}{2} g^{\mu\nu}\partial_\mu \phi \partial_\nu\phi - V(\phi)\right] \nonumber \\
+ \sum_i \int d^4 x \sqrt{-\tilde g} \left[ - \frac{1}{2} \tilde{g}^{\mu\nu}\partial_\mu \chi_i \partial_\nu\chi_i - U(\chi_i)\right].
\end{eqnarray}
Writing the action fully in terms of the metric $g_{\mu\nu}$, we find that
\begin{eqnarray}
&{\cal S}& = \int d^4 x \sqrt{-g} \left[ \frac{M_{\rm Pl}^2}{2} R - \frac{1}{2} g^{\mu\nu}\partial_\mu \phi \partial_\nu\phi \right. \nonumber \\
&-& \left.  \left( C(\phi) \sum_i \left( \frac{1}{2} g^{\mu\nu}\partial_\mu \chi_i \partial_\nu\chi_i  \right) - C^2(\phi) U(\chi_i) \right) \right],
\end{eqnarray}
where we have neglected the dark energy potential $V(\phi)$ since it plays no role in the very early universe. Theories of this kind have been studied for one field $\chi$ in e.g. \cite{DiMarco:2002eb}. The equations of motion for the fields are given by
\begin{eqnarray}
\ddot \phi + 3H\dot\phi + \frac{\partial \tilde U}{\partial \phi} = \beta C(\phi) \sum_i \dot\chi_i^2 &=& \frac{1}{2}\frac{dC}{d\phi}\sum_i \dot\chi_i^2, \\
\ddot \chi_i + (3H + 2\beta\dot\phi)\dot\chi_i + \frac{1}{C(\phi)}\frac{\partial \tilde U}{\partial \chi_i}  &=& 0.
\end{eqnarray}

We want to show that field $\phi$ is driven very quickly to the minimum of the coupling function during inflation and it is sufficient to show this for one field as the addition of additional scalar fields in the action above drive the field even faster to the minimum. In the case of one inflaton field $\chi$, the equations of motion for $\chi$ and $\phi$ are 
\begin{eqnarray}
\ddot \phi + 3H\dot\phi + \frac{\partial \tilde U}{\partial \phi} = \beta C(\phi)\dot\chi^2, \label{inf1}\\
\ddot \chi + (3H + 2\beta\dot\phi)\dot\chi + \frac{1}{C(\phi)}\frac{\partial \tilde U}{\partial \chi} = 0 \label{inf2},
\end{eqnarray}
where we have defined $\tilde U(\phi,\chi) = C^2(\phi)U(\chi)$. Noting that $\partial \tilde U/ \partial \phi = 2 C' C U(\chi)= 4 \beta \tilde U/M_{\rm Pl}$, we can see from the equation of motion for $\phi$, that it is driven by the derivative of the coupling function. If this function has a minimum, the field will be driven towards it and eventually settle at this point\footnote{For the coupling function in question, we have $\beta\approx \alpha (\phi - \phi_*)/2M_{\rm Pl}$.}. To be concrete, we consider a model with a plateau potential (see e.g. \cite{Kehagias:2018uem}), which is a model consistent with the swampland conjectures. We choose 
\begin{equation}\label{infexample}
U(\chi) = \frac{U_0}{2} \tanh^2 \frac{b\chi}{2M_{\rm Pl}}.
\end{equation}
For this potential we solve the equations of motion (\ref{inf1}) and (\ref{inf2}), to show that $\phi$ settles very quickly in the minimum of the coupling function. The minimum of $C(\phi)$ is again to be chosen at $\phi_* = 1 $M$_{\rm Pl}$ and we displace the field from the minimum by a maximal amount consistent with the swampland distance conjecture and set $\phi_{\rm ini} = 2$M$_{\rm Pl}$. We choose $\chi_{\rm ini} = 0.95$M$_{\rm Pl}$. The result for the evolution of the fields is shown in Figure 5 for the first 10 e--folds (in that figure we have chosen $\alpha =5$). As it can be seen, the DE field $\phi$ indeed settles within the first couple of e--folds to the value $\phi_*$, as expected. After that, $\phi$ is a spectator field during inflation and inflation is purely driven by the field $\chi$. In the case of several inflaton fields, the same will happen: $\phi$ will settle quickly at the minimum and inflation is driven by the fields $\chi_i$. 

The field $\phi$ will move away from $\phi_*$ as soon as the potential $V(\phi)$ becomes dynamically important in the late universe at a redshift $z\approx 1$. The scenario above requires that $\phi$ is coupled to at least one inflaton field as in the action (\ref{inflaaction}). But, as just seen, it is a rather efficient attractor mechanism. 

\begin{figure}
\includegraphics[scale=0.67]{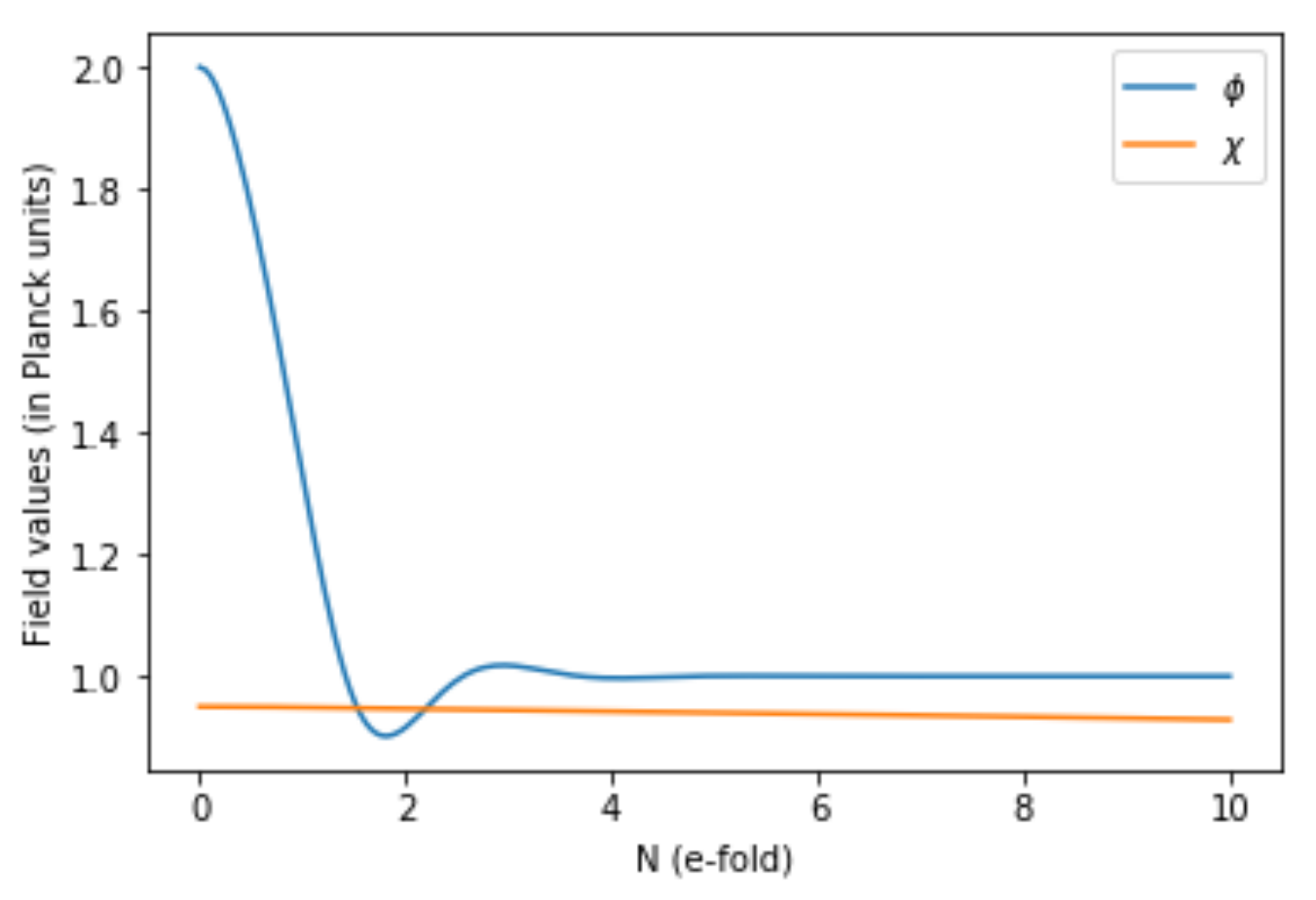}
\caption{The evolution of the inflaton field $\chi$ and the DE field $\phi$ during the first 10 e--folds during inflation, for the potential $U(\chi)$ given by eq. (\ref{infexample}). In this example, we have chosen $b=10$ and $U_0 = 10^{-5}M_{\rm Pl}^4$ for the parameter in the potential and $\alpha = 5$ in the coupling function. Inflation lasts more than 50 e--folds. The field $\phi$ is driven quickly to the minimum value $\phi_*$, whereas $\chi$ rolls down its potential.}
\end{figure}

Let us turn to a second mechanism which drives the field towards the minimum after inflation. In the very early universe, DM is initially non--relativistic and becomes in many scenarios non--relativistic before BBN at temperatures above MeV. In the case of DM with a general equation of state ($p_{\rm DM}$ is the pressure of the DM fluid) $w_c = p_{\rm DM}/\rho_{\rm DM}$, the equations governing the energy density and the scalar field become 
\be
\dot \rho_c + 3 H \rho_c(1 + w_c) = M_{\rm Pl}^{-1} \beta \dot\phi \rho_c(1-3w_c),
\ee
and 
\be\label{KG}
\ddot \phi + 3H\dot\phi + V_{,\phi} = - M_{\rm Pl}^{-1}\beta \rho_c(1-3w_c),
\ee
since the conformally coupled scalar field couples to the trace of the DM energy momentum tensor. Initially, DM is relativistic and the trace of the energy--momentum tensor (nearly) vanishes. Assuming that the potential energy does not play a role, the field is heavily damped by the Hubble expansion. At a temperature $T\sim m_{\rm DM}$ ($m_{\rm DM}$ is the mass of the DM particle), DM becomes quickly non--relativistic (which happens well before BBN for heavy DM particles) and the right--hand-side (RHS) of the Klein--Gordon equation does no longer vanish if the field is not at the minimum $\phi_*$. The field gets a 'kick' and is then driven by the term on the RHS towards the minimum value of the function $\beta$ (and hence $C(\phi)$). Such a mechanism has been studied in \cite {Damour:1993id} for scalar--tensor theories and in \cite{Brax:2004qh} for chameleon theories. In scalar-tensor theories usually all matter species are universally coupled to the scalar degree of freedom and the field gets a kick whenever a species becomes non--relativistic. In the present theory, the field gets a kick only once if there is only one species of DM. This attractor mechanism is therefore not as efficient as in a universally coupled scalar--tensor theory. To show that the kick will displace the field by at most $1 M_{\rm Pl}$, following \cite{Brax:2004qh} (Appendix 2) we can estimate the displacement by approximating the RHS of the Klein--Gordon equation (\ref{KG}) by a delta--function source, i.e. 

\be
\ddot \phi + 3H\dot\phi + V_{,\phi} \approx - \beta \frac{g_{\rm DM}}{g_*} H M_{\rm Pl} \delta(t - t_0),
\ee
where $g_{\rm DM}$ is the number of internal degrees of freedom, $g_*$ is the effective number of relativistic degrees of freedom and $t_0$ is the time of the kick at which $T\sim m_{\rm DM}$. Integrating this equation one finds the displacement \cite{Brax:2004qh} 

\be\label{kick}
\frac{\Delta \phi}{M_{\rm Pl}} \approx -\frac{g_{\rm DM}}{g_*} \beta_{t_0},
\ee
where the coupling function $\beta$ is evaluated at the time $t_0$. The expression for $\beta$ is 

\be
\beta = \frac{ \frac{\alpha}{2}\frac{\phi-\phi_*}{M_{\rm Pl}}}{1 + \frac{\alpha}{2}\left( \frac{\phi-\phi_*}{M_{\rm Pl}} \right)^2 }~.
\ee

For the extreme case that $\phi$ is displaced from the minimum by one Planck unit ($\phi - \phi_* = 1 M_{\rm Pl}$) at the time of the kick, we find that $|\beta| \leq 1$ for all values of $\alpha$. For a heavy DM particle with $m_{\rm DM} \approx 10^2~$GeV, we have $g_{\rm DM}/g_* \approx 10^{-2}$. For a somewhat lighter DM particle, $g_*$ might be somewhat smaller than 100 at the time when the particle becomes non--relativistic and the ratio $g_{\rm DM}/g_*$ may be pushed to be of order 0.1, but it is still well below one. Thus, the field is displaced by the kick by an amount much smaller than the Planck mass. If there were several DM species, the field gets a kick whenever a DM species becomes non--relativistic. The total displacement can be estimated from (\ref{kick}) by summing up individual contributions. 

To conclude, both during inflation and at the time when DM becomes non-relativistic the field is naturally driven towards the minimum of the coupling function $C(\phi)$. This attractor mechanism is a rather attractive feature of the model just discussed.

\section{Derivative couplings}
Among the extensions of the theories discussed above are theories in which derivative couplings between DM and DE are allowed. A well motivated example of such a theory is in which the conformal transformation discussed above is extended to include a disformal term. That is, instead of a purely conformal relation between the metric $g$ and ${\tilde g}$, we allow for a disformal relation of the form (the comma represents the derivative with respect to the coordinates) 
\be
{\tilde g}_{\mu\nu} = C(\phi) g_{\mu\nu} + D(\phi)\phi_{,\mu}\phi_{,\nu},
\ee
where the term containing the function $D(\phi)$ is the disformal term. Such theories appear naturally in theories with branes, in which the disformal term originates from the induced metric on the brane containing 
the DM particles. In general, the disformal coupling $D(\phi)$ introduces a new mass scale into this theory. A specific, string-inspired set-up has been studied in \cite{Koivisto:2013fta}, in which the 
functions $C$ and $D$ are of power-law form, dictated by the geometry of the higher-dimensional space. In \cite{Zumalacarregui:2012us,vandeBruck:2015ida,vandeBruck:2016hpz,Mifsud:2017fsy,vandeBruck:2017idm}, the exponential form $D(\phi) = M^{-4} \exp(2\gamma \phi)$ has been studied in detail. 

Assuming a canonical kinetic term for the DE field, the effective coupling is now specified by the function 
\be
Q=\frac{C_{,\phi}}{2C}T_{\rm DM}+\frac{D_{,\phi}}{2C}T_{\rm DM}^{\mu\nu}\nabla_\mu\phi\nabla_\nu\phi-\nabla_\mu\left[\frac{D}{C}T^{\mu\nu}_{\rm DM}\nabla_\nu\phi\right],
\ee
where $T^{\mu\nu}_{\rm DM}$ is the DM energy tensor and $T_{\rm DM}$ its trace. In cosmology, $Q$ can be written as 
\be
\beta= -\frac{Q}{\rho_c} = M_{\rm Pl} \frac{C_{,\phi} + D_{,\phi}{\dot\phi}^2 - 2D\left(\frac{C_{,\phi}}{C}{\dot\phi}^2 + V_{,\phi} + 3H\dot\phi\right)}{2\left[C + D\left(\rho_c - {\dot\phi}^2\right)\right]} 
\ee
for a pressureless fluid. For the exponential potential and exponential couplings, it was shown in \cite{vandeBruck:2015ida} that the coupling is very small in the early universe and until late in the matter dominated epoch, due to the suppression by the 
denominator. At a redshift $z\approx 3$, the coupling grows and DM particles begin to feel the force mediated by $\phi$. In Figure 6 we show the evolution of the effective gravitational constant in a purely disformal model, in which $C(\phi)=1$ and $D = 1/M^4$ and an exponential potential. The choices shown are $M=2.5$ meV and $\lambda = 0.5$ for model D1 and $M = 4$ meV and $\lambda = 0.7$ for model D2. The choice of $\lambda$ and the mass scale $M$ affect the behaviour of the effective gravitational constant substantially but both models are in the 2$\sigma$ region of allowed parameter space \cite{vandeBruck:2017idm}. We emphasize again that a growing effective gravitational constant is a prediction of these models, providing another motivation to search for equivalence principle violations in the dark sector. 

\begin{figure}
\includegraphics[scale=0.7]{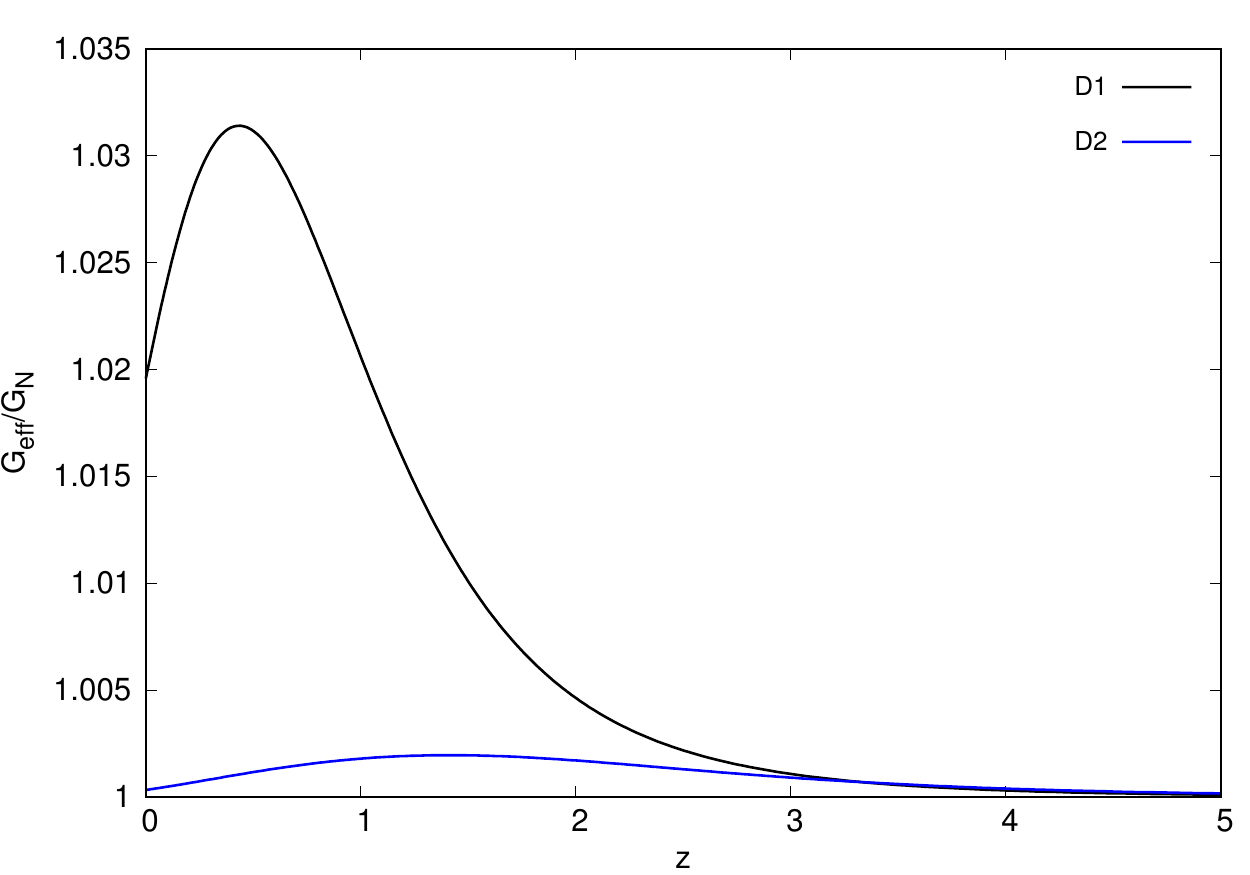}
\caption{The evolution of the effective gravitational constant for the purely disformal models D1 ($M=2.5$ meV and $\lambda = 0.5$) and D2 ($M = 4$ meV and $\lambda = 0.7$.) }
\end{figure}

\section{Conclusions}
If the swampland conjectures survive further theoretical scrutiny, they will have considerable impact on modelling of dark energy and inflation within string theory. Current observations of the universe are in excellent agreement with the $\Lambda$CDM model. In particular, the equation of state of DE inferred from observations is in very good agreement with the cosmological constant. 

In this paper we have discussed the question of DM-DE couplings in the context of the swampland conjectures. In the well--studied case of the coupling specified by (\ref{coupledquintessence}), the fifth force, which acts continuously throughout the history of the universe in this model, has to be much smaller than the gravitational force. We have argued and shown that another class of models is worth investigating, in which the fifth force between dark matter particles appears at redshifts when dark energy becomes dynamically important. We have discussed two models: one with purely conformal coupling and one which allows for disformal couplings. For the conformal coupling, we have studied the case of a coupling function which possesses a minimum. We have argued that the field is driven towards the minimum in the very early universe. In the very early universe, the DE field becomes essentially uncoupled from DM and is stuck at a fixed field value. This provides an explanation for the initial condition of the quintessence field. Deep inside the matter area, the potential energy drives the scalar field away from the minimum. Since the coupling no longer vanishes, a long--range fifth force between DM particles appears. Importantly, we have shown that since the motion of the quintessence field is slowed down by the matter coupling, larger values of $\lambda$ for the exponential potential are potentially allowed by observations. On the other hand, the enhanced gravitational constant leads to larger predicted values for $\sigma_8$ \cite{Mifsud:2017fsy,vandeBruck:2017idm}. We will study the predictions and the constraints on the parameter of the theory in detail in future work. Finally, we have pointed out that time--varying couplings between DE and DM appears naturally in theories with derivative couplings, such as in disformally coupled models. In these modes the effective coupling is a function of the DM density as well as on the time--derivative of the scalar field. 

For both types of theories, the inferred DE equation of state can mimic that of a tachyon field with $w\lesssim -1$ at intermediate redshift, but the theory predicts that, today, $w\geq-1$. It is therefore important to measure the DE equation of state for redshifts $0\leq z \leq 2$ to high accuracy. Since the effective gravitational coupling between two DM particles can today be substantially bigger than gravity alone, our work motivates searches for equivalence principle violations in the dark sector at various redshifts and in the non--linear regime \cite{Baldi:2008ay}. One test has been suggested in \cite{Kesden:2006vz}, using satellite galaxies which are tidally disrupted by the Milky Way. It would be worth investigating this test in the context of the theories discussed here. 

\noindent {\bf Acknowledgements} We are grateful to Jurgen Mifsud for providing us with the modified CLASS code \cite{Blas:2011rf} and to Sebastian Trojanowski for useful discussions. CvdB is supported (in part) by the Lancaster-Manchester-Sheffield Consortium for Fundamental Physics under STFC grant: ST/L000520/1. CCT is supported by a studentship from the School of Mathematics and Statistics at the University of Sheffield.

\bibliography{dmdebib}

\end{document}